# Numerical analysis of the propagation modes of photo-switching PDMS-arylazopyrazole optical waveguide and thin-film spectroscopic characterization

Golsa Mirbagheri*[a], Ikemefuna Uba[a], Kesete Ghebreyessus[b], Uwe Hömmerich[c], Seth Fraden[d], Michael Stehnach[d] and Demetris Geddis[a]

[a] Department of Electrical and Computer Engineering, Hampton University, Hampton, VA, 23668, USA; [b]Department of Chemistry and Biochemistry, Hampton University, Hampton, VA, 23668, USA; [c]Department of Physics, Hampton University, Hampton, VA, 23668, USA; [d]Physics Department, Brandeis University, Waltham, Massachusetts 02453, USA

## ABSTRACT

A new light responsive arylazopyrazole (AAP) containing polymer matrix thin film is fabricated by spin-coating of different concentrations of the AAP azo dye into the polydimethylsiloxane (PDMS) polymer at 150°C. The new AAP molecular switch was also used to fabricate a solid-state PDMS-AAP waveguide by contact lithography and soft replica modeling methods in the micrometer scale. The refractive index of the spin-coated photoswitchable material can be modulated via the reversible *trans-to-cis* photoisomerization behavior of the AAP unit using different concentrations. When 0.01 M solution of the AAP unit was used, the refractive of the composite was 2.32 in the trans state and dropped to 1.85 in the cis state in the operating wavelength of 340 nm. At higher concentrations of 0.020 and 0.03 M, a wide refractive index tuning is achieved under the same wavelength. In 0.030 M the refractive index was 2.65 for the trans state and 2.0 for the cis state. The results suggest that the increase in refractive index tuning is related to the concentration of the AAP unit of the composite film. Theoretically, the spectral properties of the composite film are also simulated with two methods: 1) the Maxwell Equations; and 2) the frequency dependent finite element, showing excellent agreement for the different propagation modes of the proposed waveguide for regulated signals of 365/525 nm wavelengths. Furthermore, the photoisomerization of the PDMS-AAP thin film is analyzed with UV-vis spectroscopy to demonstrate the isomerization responses of the AAP moiety in the solid state. Additionally, preliminary photomechanical actuation properties of the composite film have been investigated. The PDMS-AAP waveguide described in this study provides a new approach for optically tunable photonics applications in the Visible-IR region.

## 1. INTRODUCTION

Photoresponsive smart materials whose properties can be reversibly switched between two distinct states by light stimulus have been widely developed in diverse optochemical platforms [1, 2]. Previous work on photoresponsive materials that undergo light-induced reversible *trans-to-cis* conformational changes has mostly been concentrated on the highly versatile azobenzene photoswtiches. Azobenzenes undergo photo-induced *trans-to-cis* isomerization upon exposure to light of specific wavelengths. Despite their remarkable success, azobenzenes often suffer from two performance issues: incomplete switching or low thermal stability of the metastable *cis*-isomer [2, 3, 4, 5, 6]. Arylazopyrazoles (AAPs) have been recently introduced as superior analogues of azobenzenes characterized by near quantitative conversion between two states and superior thermal half-life times of the metastable state [7, 8, 9, 10]. AAPs undergo efficient reversible *trans-to-cis* photoisomerization upon alternating irradiation with UV or green light to achieve a conformational change. The facile synthetic access and the superior optical properties of AAPs make them a favorable choice in the fabrication of guided optics, optoelectronics, and optofluidic platforms [9, 10].

*golsa.mirbagheri@hamptonu.edu

The change in molecular configuration leads to a significant change in electronic and steric structure, and thus allows for switching of intrinsic material properties. One of the intrinsic material properties addressed in this study is refractive index. The photo-chemical modulation of the refractive index in polymeric materials and their composites is of significant interest for optical applications such as polymeric waveguides, optical switches and optical data storage devices. Therefore, due to the important role of the refractive index in polymeric material, photo-chemically induced index modification in photonic devices gained considerable attention [11]. Tuning the propagation modes of the waveguide modifies the strength of the interaction force between the waveguide channels [10, 12]. One way to inspect this tuning is through the embedding of photochromic organic molecules into the polymeric waveguides. However, while there are many examples of photoresponsive waveguides formed by an azobenzene based polymers that can achieve refractive index modulation upon illumination with specific wavelength of light [1, 2], to the best of our knowledge, there are only few reported studies of polymeric waveguides based on the novel AAP molecular switches [10, 11]. Hence, the present study aims to exploit the novel properties of the AAP molecular switches for the development of new photoswitchable light responsive polymeric waveguides that would enable the reversible photomodulation of the refractive index of azo polymer composite films.

We have previously demonstrated the fabrication of PDMS-AAP composite thin films and studied the refractive index changes through the *trans-to-cis* isomerization of the AAP unit embedded in the polymer matrix [6, 9, 10]. The PDMS-AAP matrix exhibits 100% light-triggered reversible efficiency, with maxima refractive index change during isomerization process [10, 13, 14]. In this work, we have extended the study through the design of new AAP based molecular switch and its use for the fabrication of PDMS-AAP based waveguide, where the index of PDMS-AAP composite core is adjusted *in-situ* by alternating exposure to UV and green light. The immersed core simulated in section 2 is surrounded by flexible PDMS clad which is transparent in the UV-visible region. Our previous study demonstrated the feasibility by closed-form analysis of Maxwell equations and illustrated the electric field modes of the guided light in the PDMS-AAP core [10]. In section 2, the fabrication of the PDMS-AAP composite film is carried out by spin-coating of the different concentrations of the AAP and PDMS units and baked at 150°C. Then, we outline the fabrication of the PDMS replica mold with soft lithography. The contact photolithography method described in this study is used to make the PDMS grooves in a micrometer scale. This initial study is undertaken with the goal of expanding the soft lithography technique for the fabrication of nanoimprint lithography to create nano-grooves of the PDMS [15, 16]. In section 3, the simulation of the waveguide is performed by illustrating the electric field modes of the guided light into the PDMS-AAP core through numerical finite element method [17, 18]. The photoisomerization behavior of the photochromic molecules embedded in the composite thin film characterized by UV-vis spectroscopy through alternating exposure of the core to 365 nm and 525 nm wavelengths as described in section 4. Additionally, the mechanical properties of the composite thin film are investigated by regulating the UV and green lights, resulting in a slight nanoscale bending actuation due to isomerization of the AAP molecules embedded in the polymer matrix. Tunable subwavelength materials enable changing the light-matter interactions could be used in novel applications including sensing and imaging applications [19, 20, 21].

## 2. EXPERIMENTAL SECTION

### A. Material
All reagents were commercial grade and used as received without further purification.

### B. Preparation of the composite
Commercial Sylgard 184 PDMS kit from Dow Corning Inc was mixed in the standard 10:1 ratio using $1.0 \pm 0.04$g elastomer and $0.1 \pm 0.02$g curing agent. Appropriate amounts of the AAP were dissolved in acetone to make 0.01M, 0.02M and 0.03M solutions. 50μL of each molar solution were added to dedicated PDMS mixtures, stirred manually with glass rod and then deposited on previously cleaned glass substrates via spin-coating at 1000 rpm. The high spin speed eliminated the need to degas the blend prior to coating. The coated substrates were then cured at 150°C for 20 minutes. The resulting PDMS-AAP photo composites were transparent and could easily peel off form the substrates.

## C. Surface characterization

Surface morphology of the composites was investigated with JEOL 6390 scanning electron microscope for structuring or clusters that would indicate phase segregation. A snippet of each composite was cut and peeled off for the imaging, which was done at (i) 1kV/180x magnification and (ii) 13kV/95x magnification.

## D. UV-Vis Spectroscopy

Optical properties of the composites were determined via absorbance spectra recorded with Shimazu 3600 spectrophotometer in UV–vis range and compared to that of plain PDMS film of same thickness. A free space baseline was set in 270–800 nm wavelength range to eliminate stray interference and ensure that only absorbance of composites and PDMS films were recorded. A snippet of each sample was peeled off and mounted on sample holder for measurements. First, absorbance spectrum of the plain PDMS was acquired as a reference to be used to assess deviations in the composites because of the integrated photoswitch. Measurement on the composites was done in two stages: (i) before UV irradiation, when embedded molecules were in the *trans* state and (ii) after 5minutes UV (λ = 365 nm) irradiation when molecules were in cis state. We tagged these stages pristine and UV, respectively. The 5 minutes exposure time was informed by prior observation that our AAPs achieve photo-stationary state (maximum trans-cis conformational change) at this duration [11]. The refractive indices of the plain PDMS and the composites were then determined from their absorbance spectra.

## 2.1. Fabrication of the PDMS-AAP composite thin film via Spin-Coating Method

The AAP molecular switch referred was synthesized using commercially available reagents as previous described by our group [9]. Preparation of the PDMD-AAP matrix was carried out using polydimethylsiloxane (elastomer and curer) of the Sylgard 184 silicone elastomer kit and AAP solution via spin-coating method. First, 0.10 g of the AAP molecular switch was dissolved in 10 mL of acetone to make 0.025 M solution at room temperature, as shown in Figure 1b.

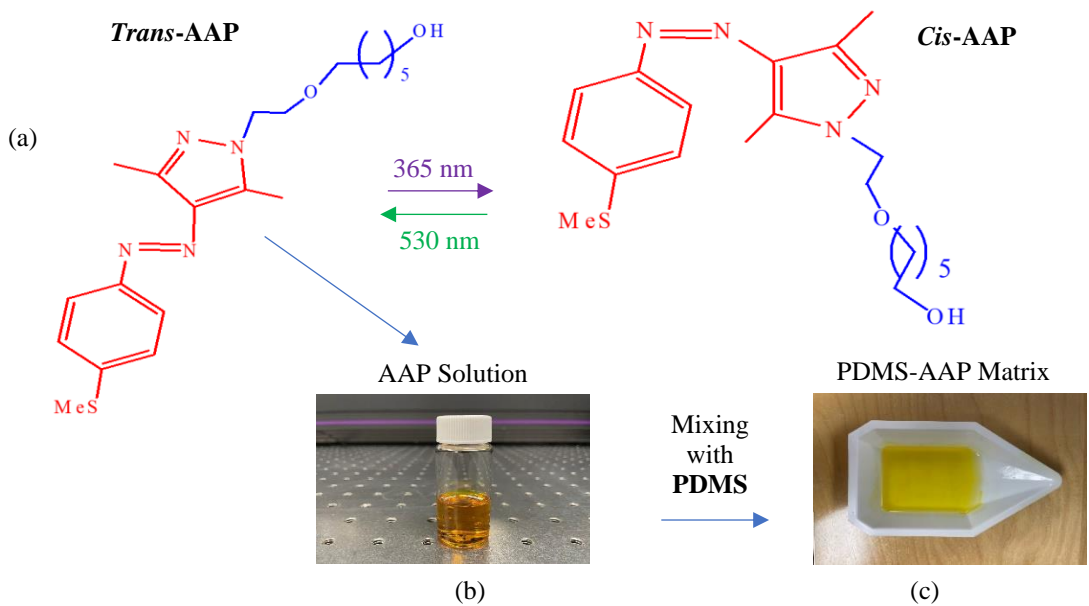

Figure 1. (a) Chemical structure and photoswitching behavior of the AAP moiety; and (b-c) fabrication process of the PDMS-AAP matrix.

Next, the polydimethylsiloxane elastomer and curer are stirred in a 10:1 weight ratio. Then, 1.0 mL of the 0.025 M solution of the AAP was poured into 20 g of PDMS matrix and stirred for few minutes to make a homogenous clear gel as shown in Figure 1 c. The uniform gel is spin-coated on the microscope slide at a high-speed of 2500 rpm for 2 minutes to make 75 μm thin film. Then, the coated gel is baked at 150°C on a hotplate for 20 minutes. This procedure was repeated with different concentrations of the AAP photoswitch to fabricate different PDMS-AAP composite thin films to probe the effect of the concentration of the AAP photoswitch on the isomerization process and thickness of photochromic film.

## 2.2 Fabrication of the PDMS-AAP Waveguide via the Photolithography Process

### 2.2.1. Design of the Primary Mold

The grooves of the PDMS clad are designed and printed on the mylar mask. As shown in Figure 2a, each square of the nine on the mold mask contains multiple rectangular shaped channels with different dimensions and gaps in micrometer scale. The mask was designed with CAD software and the mylar mask was made by Artnet Pro, as shown in Figure 2b.

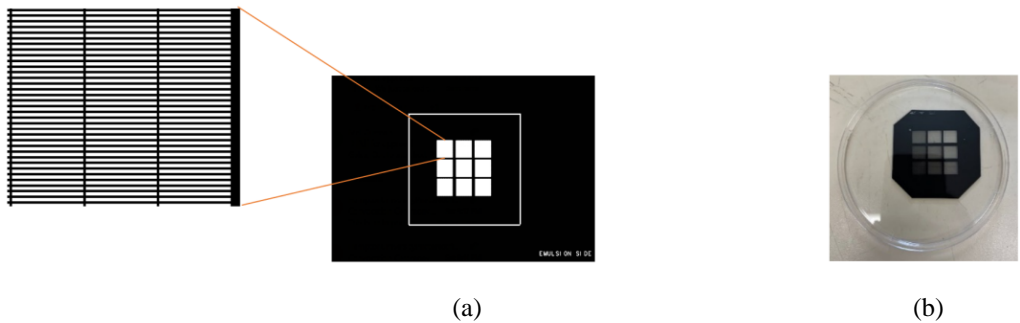

(a) (b)

Figure 2. Design of the mask in CAD (a) and primary mylar mask (b)

### 2.2.2. Making the PDMS Replica Mold

The mylar mask is used to fabricate the primary mold in contact lithography process. First, the negative resist (SU-8 3025) is spin-coated on a 3 inches wafer with 2500 rmp for 48 seconds to give a 20 μm thick film. The coated wafer is then soft-baked for 10 min at 95 °C. In the contact lithography step, the mylar mask is placed between a quartz and the photoresist-coated mold wafer to print the channels on the wafer. The wafer is exposed to UV light for 15 seconds and then baked at 95 °C for 5 minutes. In post exposure step, the wafer is developed in propylene glycol monomethyl ether acetate (SigmaAldrich 484431-4L) for 10 minutes and rinsed with isopropanol alcohol (IPA), as shown in Figure 3. Finally, the wafer is baked for 5 minutes at 150 °C after development.

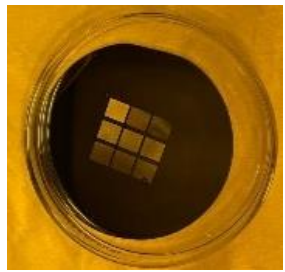

Figure 3. Primary mold wafer coated with negative photoresist in lithography room

In order to make the PDMS replica mold, 30 g Silgard is blended with 3 g curer and stirred with conditioning mixer and bubbles in the matrix are removed using a desiccator. Then, the PDMS is poured into the primary mold wafer and baked in an oven at 70 °C for 1 hour. The PDMS replica was peeled off from the mold wafer (Fig. 4a). To fabricate the photoswitch waveguide, the PDMS-AAP is poured into the groves of PDMS replica (Figure 4b). The extra PDMS-AAP matrix around the grooves was removed by blade or pallet knife after couple of hours, as shown in Figure 4c.

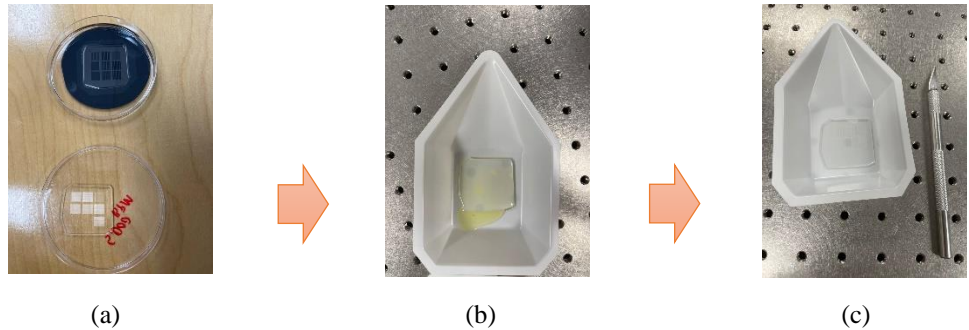

(a)                            (b)                         (c)

Figure 4. (a) PDMS replica mold peeled off from primary wafer, (b) PDMS-AAP poured into the grooves of PDMS replica, and (c) extra PDMS-AAP matrix removed around the grooves with blade.

### 2.2.3. Morphology of the PDMS-AAP composite film

The SEM images of the PDMS replica mold are displayed in Figure 5a-b. The SEM images show that the width of the canals is 27.3 µm. Figure 5c shows image of the PDMS-AAP filled replica channels. Although the fabricated PDMS features have the size of micrometer, this initial photolithography study gives a clear idea for the next phase of the project, making the nanoscale hollows of the primary and replica molds with nanoimprint lithography method.

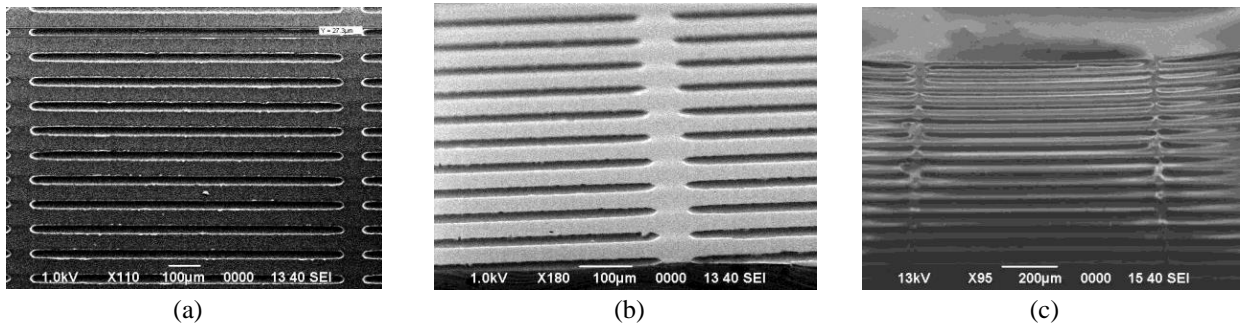

(a)                            (b)                         (c)

Figure 5. (a-b) SEM images of the PDMS replica mold, and (c) the replica channels filled with PDMS-AAP.

# 3. RESULTS AND DISCUSSION

## 3.1. Simulation of the E-Filed Profiles of the Waveguide

### 3.1.1. Wave Confinement in PDMS-AAP Waveguide with Maxwell Equations

In our previous work, we designed the buried waveguide structure composed of 320 nm thick PDMS-AAP as the core surrounded by plain PDMS. The refractive index of the embedded PDMS-AAP is maximum at the operating wavelengths of 340 nm and 450 nm during the reversible isomerization process by the UV-visible regulating sources [10, 11]. The 640 nm thick pliable PDMS clad/substrate is transparent to the UV-green regulator light, so the light reaches the immersed PDMS-AAP with the least attenuation. In this study, wave confinement was investigated with Maxwell's equations in the isomeric states of the composite core for different design properties. Based on the effective index method [10], the waveguide structure was defined in terms of normalized parameters, shown in Eq. (1), for the wave $E_y(x) = E_o(x,y)e^{i(\beta z - \omega t)}$ polarized in the y and propagating in z direction. where v is v-number (normalized frequency), b represents relative index, and $\alpha$ is asymmetry factor.

$$v\sqrt{1-b} = (m-1)\pi + tg^{-1}\sqrt{b/(1-b)} + tg^{-1}\sqrt{(b+\alpha)/(1-b)}$$

and $\quad \alpha = \frac{(n_{clad}^2 - n_{sub}^2)}{(n_{core}^2 - n_{clad}^2)}$  (1) [10]

For a specific waveguide design, the normalized frequency at operating wavelength $\lambda$, is related to the core thickness and indices, as shown in Eq. (2).

$$v = \frac{2\pi d}{\lambda}\sqrt{(n_{core}^2 - n_{clad}^2)}$$  (2) [22]

Figure 6 shows the b-v plot for the waveguide where the vertical red line marks the v-number in trans state. At the operating wavelength of 340 nm, three modes of the waveguide in the trans state are observed at the normalized frequency v = 7.0. However, in the cis state, only one mode is detected in at normalized frequency v = 2.4 for the same operating wavelength. The v-numbers are calculated in the same way for 450 nm, based on b values [10].

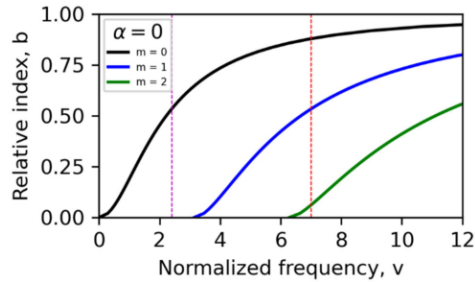

Figure 6. Relative index / normalized frequency chart for 320 nm-thick PDMS-AAP core at operating wavelengths 340 nm and 450 nm [10].

### 3.1.2. Propagation Modes Extraction of PDMS-AAP Waveguide Using Ansys HFSS

In this work, the photoswitchable waveguide is simulated with finite element method, as shown in Figure 7. The PDMS-AAP core has a dimension of 320 nm, surrounded by 160 nm plain PDMS for each side of the cube. The overall PDMS thickness and length is $640 \times 640$ nm. The excitation ports were designed with floquet ports on top and bottom of the vacuum box in z-direction, while primary/secondary boundary conditions were considered for four sides of the vacuum rectangular cuboid. The number of modes can be adjusted in the floquet ports.

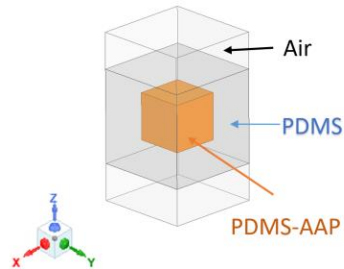

Figure 7. Photoswitch waveguide structure designed with Ansys HFSS.

In our previous work, we studied the reversible refractive index modulation of a photoswitchable PDMS-AAP based composite film. As illustrated in Figure 8, upon irradiation with light of specific wavelength, the AAP unit embedded in the PDMS-AAP matrix underwent reversible *trans-to-cis* isomerization, which was manifested by a drop in absorbance of the maximum absorption band at ~ 340 nm and by decrease in the refractive index. The measured refractive index of the embedded AAP core is used to compare the two numerical methods (Maxwell and finite element) for both the cis and trans states upon alternating illumination with UV ($\lambda = 365$ nm) and green ($\lambda = 525$ nm) light, respectively. Figure 8 shows the refractive index values of the PDMS-AAP composite from our previous study [11]. At the operating wavelength of 340 nm, the refractive index reached to maximum 2.0 in the trans state. In the cis state, the core index at operating wavelength of 450 nm is 1.65. However, the refractive index of the organic core doesn't change significantly beyond 450 nm for trans and cis states. The PDMS index doesn't have notable variation at operating wavelengths 340 and 450 nm for both isomeric states.

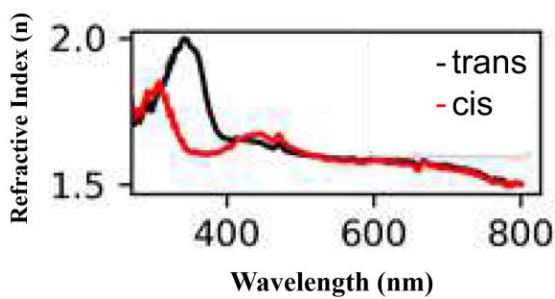

Figure 8. Refractive indices of PDMS-AAP core measured for trans and cis [11]

In the present study, the refractive index of the PDMS-AAP composite fabricated using a new photoswitch was measured based on the absorption spectra of the film using different molar concentrations of the AAP unit for both the trans and cis states (Figure 9) [11]. In the trans state (Figure 9a), the refractive index maxima of the 0.01 M composite is 2.32 at 327 nm, whereas in the cis state the index of the film drops to 1.85. On the other hand, the maximum index for the cis state was 1.72 at 450 nm wavelength. Notably, as significant increase in the refractive index change was observed when the concentration increased to 0.020 and 0.03 M. Refractive indices of the 0.02 and 0.03 M solutions of the AAP in the PDMS-AAP composite were 2.65 in the trans state and 2.1 in the state at an operating wavelength of 326 nm (Figure 9b-c). These show that refractive index can be effectively tuned by changing the concentration of the azo dye embedded in the polymeric composite film. Additionally, these results indicate a remarkable enhancement of the refractive index change compared to our previous study [10]. We also presume that the different structure of the newly designed AAP may have played a role in enhancing the refractive index changes.

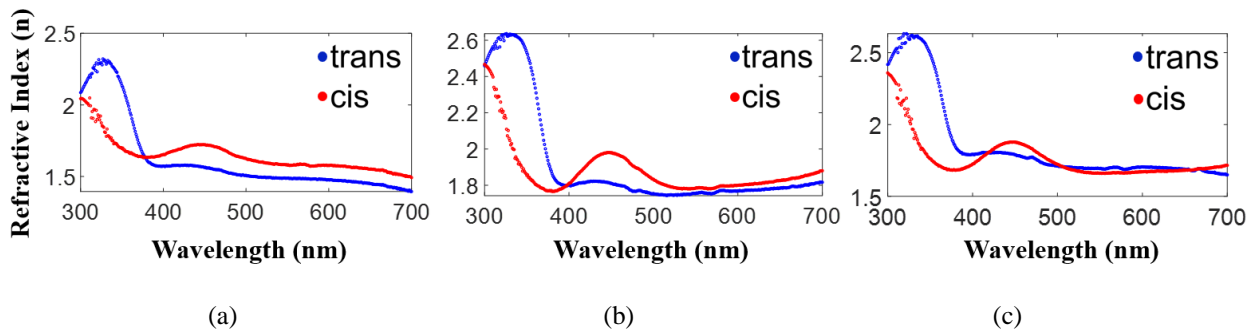

(a)                                  (b)                                  (c)

Figure 9. Refractive index of three different molarity concentrations (a) 0.01 M, (b) 0.02 M and (c) 0.03 M of the PDMS-AAP composite thin film

The propagation modes of the optical waveguides obtained based on the frequency vs imaginary part of the gamma are illustrated in Figure 10. At frequencies higher than 656 THz, three modes of the PDMS-AAP based waveguide are observed in the trans state. Similar to the effective index method, we see only one mode in the cis state at frequencies lower than 470 THz. It should be mentioned that based on Figure 8,the refractive index of the waveguide core is 1.63 at 450 nm in the trans state, which implies that we could see only single mode at lower frequencies.

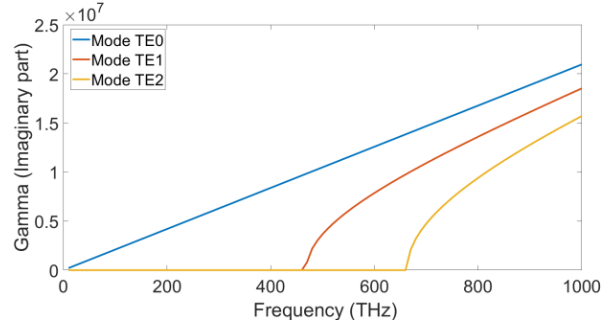

Figure 10. HFSS Simulation of wave propagations of optical waveguide in cis and trans isomeric states

The relation between the imaginary part of the gamma in the finite element method and relative index b in the effective index method is explained in Eq. (2). The propagation modes of the photoswitch structure show excellent agreement for the two methods of Maxwell Equation in Figure 6 and frequency dependent finite element method described in Figure 10.

$$E_y = E_0(x, y) \, e^{i\beta z} = E_0(x, y) \, e^{\gamma z} \text{ and } \beta = n_{eff} \, k_0$$

then $\gamma = i\beta = i \, n_{eff} \, k_0$   where   $b = \frac{n_{eff}^2 - n_{clad}^2}{(n_{core}^2 - n_{clad}^2)}$   (2) [10]

The transverse e-field profiles in the core are simulated on excitation ports of the structure in $\hat{z}$ direction. Three fundamental modes (TE0, TE1 and TE2) of the waveguide in the trans state are observed on the PDMS-AAP core at frequency of 700 THz as shown in Figure 11a. The waveguide core in the cis isomeric state simulated at an operating wavelength of 340 nm shows single mode (Figure 11b). In the cis state, all higher modes are suppressed at lower frequency of 240 THz due to the resulting refractive index change by photoisomerization process. At an operating wavelength 340 nm, the refractive index of the core reaches to 2 and 1.65 for trans and cis states respectively, as shown in Figure 8.

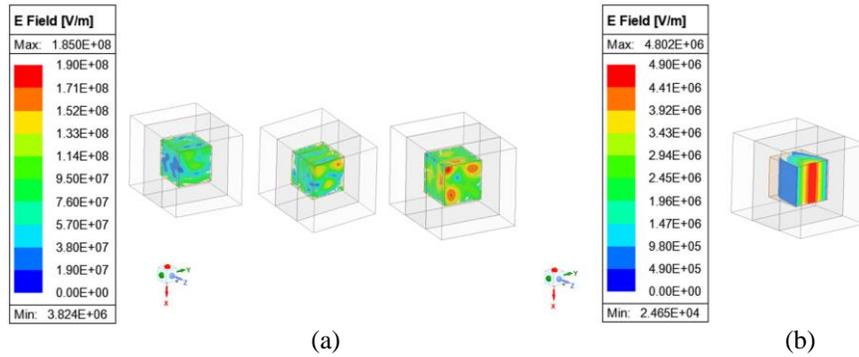

(a)                                        (b)

Figure 11. Transverse electric field profiles of the PDMS–AAP buried waveguide at an operating wavelength of 340 nm in the (a) trans and (b) cis states analyzed by HFSS.

Furthermore, the transverse electric field modes of the photoswitch core are analyzed in the visible region as described in Figure 12. At an operating wavelength of 450 nm, the PDMS-AAP buried core supports only one mode at frequencies lower than 470 THz. Interestingly, the number of modes can be increased in the cis state by changing the thickness of the photoswitch core. At wavelength 450 nm, the refractive index of the core is 1.63 in pristine and 1.67 in UV light, as described in Figure 8 above. For comparison purpose, TE profiles of the PDMS-AAP composite are simulated at 160 THz and 200 THz in the trans and cis, respectively. The results show that there is good agreement with the finite element method (Figure 12a-b), and the closed-form analysis of the waveguide in Figure 12c (6). Thus, the independent finite-element simulation re-affirms waveguiding ability of the composite and validates the closed-form analysis described in our previous report.

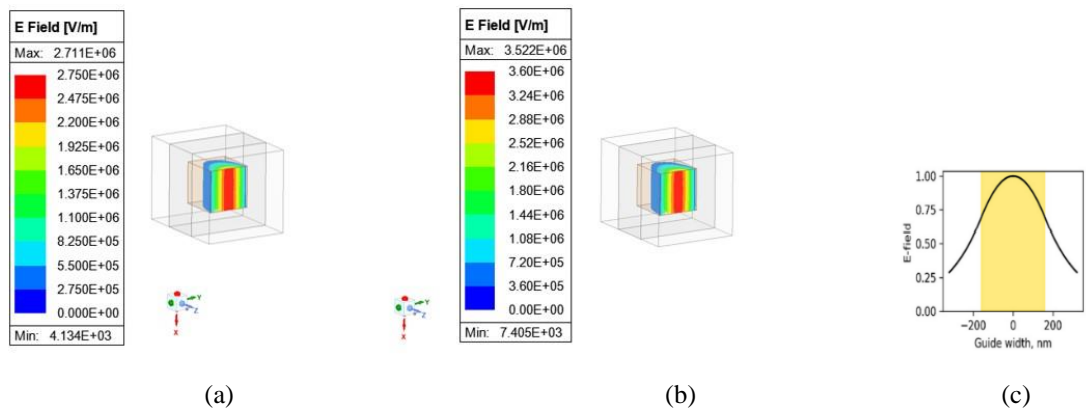

(a)                                  (b)                                  (c)

Figure 12. Transverse electric field profiles in PDMS–AAP buried waveguide at wavelength 450 nm in trans (a) and cis (b) states by HFSS. (c) Electric field profile in the cis state generated via closed-form analysis of Maxwell Equations; shaded region is the PDMS-AAP core [10]

### 3.1.3. Photochromic Properties of the Buried Waveguide Structure

The photochromic characteristic of the AAP molecular switch was examined in the UV-visible region. The color of the AAP-acetone solution changes to dark orange after irradiation with UV light for 5 minutes, as shown in Figure 13b. The same solution changes its color back to bright yellow as pristine upon irradiation with 530 nm (Figure 13a), due to the reversible isomerization of AAP molecular switch.

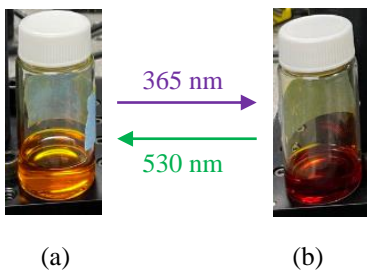

(a)                  (b)

Figure 13. Photochromic changes of AAP molecular switch dissolved in acetone upon irradiation of (a) UV ($\lambda$= 365 nm and (b) green ($\lambda$ = 530 nm) lights.

## 4. PHOTOISOMERIZATION SPECTRAL CHARACTERIZATION

### 4.1. UV-vis Spectral Characterization of the PDMS-AAP composite film

The photoisomerization responses of the PDMS-AAP composite thin film in the solid-state was characterized by UV-vis spectroscopy. The changes in the UV–vis absorption spectra of the AAP unit embedded in the composite film as a function of time is depicted in Figure 14. As illustrated in Figure 14, the UV-vis absorption spectra of the PDMS-AAP composite film displays characteristic picks of the *trans-to-cis* isomerization of the embedded AAP unit upon irradiation with UV light at λ = 365 nm and green light at λ = 525 nm, respectively. The absorbance spectra of the composite film show intense band at 340 nm and weak bands at 440-450 nm in the visible region, which is characteristic features of the *trans*-isomer. Upon irradiation with UV light (λ = 365 nm) for 2-5 minutes, the trans-form of the AAP unit embedded in the polymer matrix was easily converted to the *cis*-isomer, resulting in new absorption bands at 290 nm and 450 nm. This was also clearly shown by significant decrease in the absorbance of the maxima at 340 nm and increase of the 450 nm band, which is indicative of the formation of the *cis*-isomer. This observation shows both the relatively fast *trans-to-cis* isomerization of the AAP molecular switches embedded in the polymer matrix and tunability of the optical properties of the composites.

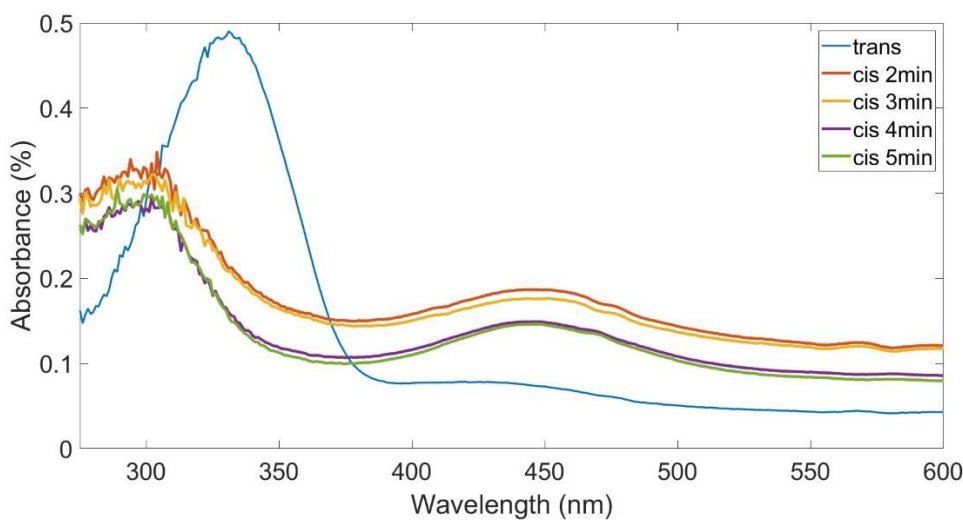

Figure 14. Experimental UV-vis Spectroscopy of PDMS-AAP.

### 4.2. Mechanical Characterization of the PDMS-AAP Thin Films

To study the photomechanical behavior, a thin slice of the PDMS-AAP composite film is placed on the needle as shown in Figure 15. The photomechanical motion/actuation of the light-responsive solid-state thin film was then monitored with a microscope. The results show that the slice thin film bend ~ 200 nm away from the light source after radiation with UV light (λ = 365 nm) for 5 minutes (Figure 15). The distance of the 45 $\mu$m thick layer from the needle is measured with Gwyddion software in Table 1.

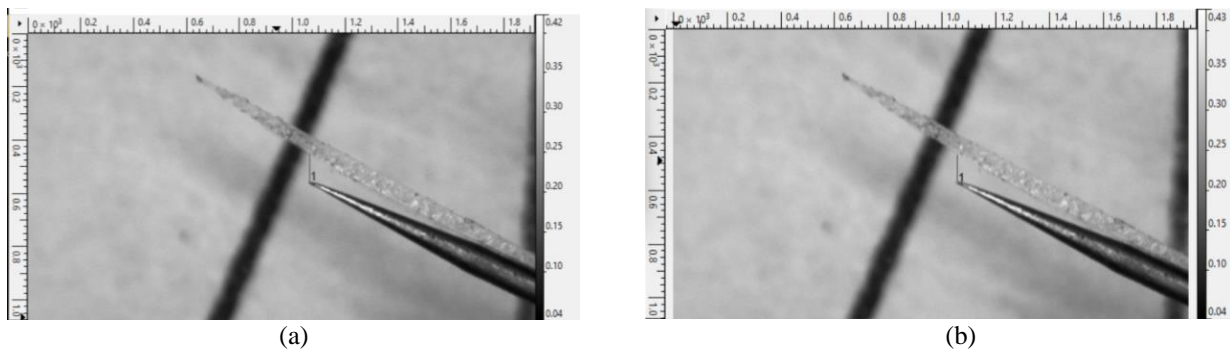

Figure 15. Microscope images of the thin photochromic layer on the needle in trans (a) and cis (b) state.

Table 1: Measuring the film by alternative exposure of UV and green light at room temperature

| Irradiation | Distance | Remark |
|---|---|---|
| Green light (*trans*) | $26.59\mu m$ | Initial width |
| UV light (*cis*) | $26.85\ \mu m$ | Away from needle |

## 5. CONCLUSION

In this work, we have designed and theoretically simulated a light-responsive waveguide that exhibit efficient reversible *trans-to-cis* photoisomerization an arylazopyrazole (AAP) unit embedded in a polydimethylsiloxane (PDMS) polymer matrix. The PDMS-AAP composite film was fabricated by mixing the photochromic AAP moiety into the flexible PDMS matrix via spin-coating. Additionally, a solid-state waveguide was fabricated by soft lithography and replica modeling methods with the aim to construct nanometer scale channels. The propagation modes of the composite film were also simulated with finite element method, tuned by the refractive indices of the waveguide in isomeric states upon alternating irradiation with UV and green lights. The results were in good qualitative agreement with the closed-form refractive index analysis described in our previous work, thereby re-establishing the wave-guiding property of the composite. It has also been shown that the concentration of the AAP unit embedded in the composite film plays an important role, in which increasing concentration has been demonstrated to significantly enhance the refractive index changes. The remarkable photo-induced refractive index changes suggest the potential applicability of the PDMS composite film doped with the novel AAP azo dyes for nanophotonic optomechanical platforms.

## 6. ACKNOWLEDGEMENT

National Science Foundation (NSF) supported this work through grant NSF-DMR 1827820 for Partnership for Research and Education in Materials (PREM). We would also like to show our gratitude to Breon Mccray, Anthony Gillespie and Jihaad Barnett for assisting in constructing the UV/green lights and synthesizing the composite film.